\documentclass[prb,twocolumn,amsmath]{revtex4}
\input {epsf.sty}
\usepackage{epsfig}
\usepackage{amssymb}
\begin{document}
\title{Two-dimensional Vortices in Superconductors}
\author{Bo Chen, W. P. Halperin,}
\affiliation{Department of Physics and Astronomy,\\
      Northwestern University, Evanston, Illinois 60208}
\author{Prasenjit Guptasarma,}
\affiliation{Department of Physics,\\
      University of Wisconsin-Milwaukee, Wisconsin, 53211}
\author{D. G. Hinks,}
\affiliation{Materials Science and technology Division,\\Argonne
National Laboratory,
      Argonne, Illinois, 60439}
\author{V. F. Mitrovi{\'c},}
\affiliation{ Department of Physics, \\Brown University, Providence,
Rhode Island, 02912}
\author{A. P. Reyes, P. L. Kuhns}
\address{National High Magnetic Field Laboratory
     Tallahassee, Florida 32310}
\date{Version \today}

\begin{abstract}{\bf Superconductors have two key characteristics.  They  expel
magnetic field and they conduct electrical current
with zero resistance.  However, both properties are compromised in
high magnetic fields which can penetrate the
material and create a mixed state of quantized vortices.   The
vortices move in response to an electrical current
dissipating energy which destroys the zero resistance
state\cite{And64}.  One of the central problems for
applications of high temperature superconductivity is the
stabilization of vortices to ensure zero electrical
resistance.   We find that vortices in the anisotropic superconductor
Bi$_{2}$Sr$_{2}$CaCu$_{2}$O$_{8+\delta}$
(Bi-2212) have a phase transition from a liquid state, which is
inherently unstable, to a two-dimensional vortex
solid.  We show that at high field the transition temperature is
independent of magnetic field, as was predicted
theoretically for the melting of an ideal two-dimensional vortex
lattice\cite{Fis80,Gla91}. Our results indicate
that the stable solid phase can be reached at any field  as may be
necessary for applications involving
superconducting magnets\cite{Has04,Sca04,COHMAG}.  The vortex solid
is disordered, as suggested by previous studies
at lower fields\cite{Lee93,Cub93}. But its evolution with increasing
magnetic field displays unexpected threshold
behavior that needs further investigation.}
\end{abstract}

\maketitle

\vspace{11pt}

Shortly after the discovery of cuprate superconductivity it was
recognized that their high transition temperature
would mean that thermal fluctuations can produce a liquid vortex
state\cite{Nel89}.  In fact the thermodynamic
transition to superconductivity in a magnetic field occurs between a
thermally fluctuating liquid vortex phase to
one or more solid phases\cite{Bla94}.  The liquid vortex state is
inherently unstable with non-zero electrical
resistance.  For extremely anisotropic materials, like Bi-2212, the
liquid phase covers a wide range of
temperature;  but it is not known exactly how wide this is in high
magnetic fields. For fields perpendicular to
the conducting planes, $H||c$-axis, the transition temperature
between liquid and solid vortex phases,
$T_{m}(H)$, is principally controlled by vortex-vortex interactions
which get stronger as the density of vortices
increases, proportional to the field.  The supercurrents that form
each vortex are mainly confined to the
conducting planes and, in high field, they lose their coherence from
one plane to the next so that vortices are
expected to become two-dimensional. It is remarkable that the
theory\cite{Gla91} for the two-dimensional vortex
melting transition has only one parameter, the magnetic field
penetration depth, and that this simple picture has
not yet been experimentally confirmed.

\begin{figure}[h]
\vspace{0.3in}
\includegraphics[width=8 cm]{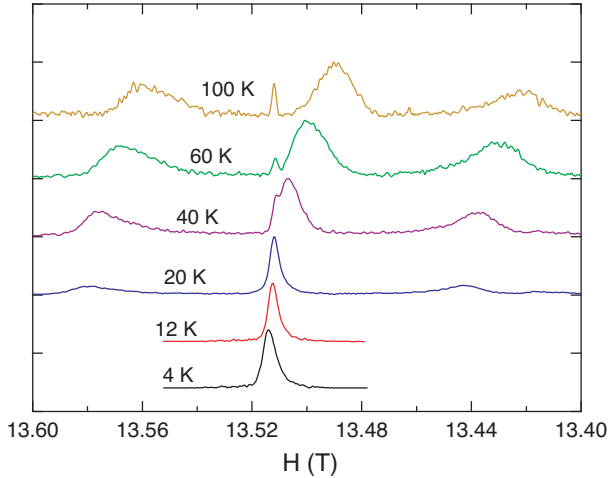}
\suppressfloats [b]
\caption[Spectra]{Spectra of $^{17}$O NMR in Bi-2212 for
magnetic field  parallel to the $c$-axis at fixed
frequency. The sample is an overdoped, 28 mg, single crystal of
Bi$_{2}$Sr$_{2}$CaCu$_{2}$O$_{8+\delta}$,
(Bi-2212), with T$_{c} = 75$ K, with
$\approx 60\%$ of
$^{16}$O exchanged for $^{17}$O. The optimally doped T$_{c} = 93$ K.
In field-sweep experiments (T = 20, 40, 60, 100 K), a
decreasing Knight shift moves the NMR spectrum to the left. At high
temperatures two oxygen sites can be distinguished.  The
central transition,
$\left\langle {-{1 \over 2}\leftrightarrow +{1 \over 2}}
\right\rangle$ for oxygen in the copper-oxygen plane,
O(1), is the wide line near 13.49 T at 100 K. The  central transition
for the oxygen in the strontium-oxygen
plane, O(2), is the narrow line near the zero Knight shift position
at 13.51 T.  The other two peaks are
quadrupolar satellites of the O(1) resonance. For T= 4 and 12 K we
show the Fourier transform of the echo at fixed field,
having a
smaller spectral bandwidth so that satellites are not observed here.
Our separate measurement of the satellite intensity at
$T/T_{m}$ = 0.67 compared to the
central transition is the same as that at $T/T_{m}$ = 1.33, confirming this
picture. The spectra at different temperatures are normalized to have the
same peak intensity of the O(1) central transition. }
\label{Fig1}
\end{figure}

We use nuclear magnetic resonance (NMR) of
$^{17}$O to detect the melting transition of vortices as a function
of temperature and magnetic field. NMR can be
performed on selected elements in site specific locations in the
structure\cite{Tro90,Tak94} as we see in the
$^{17}$O spectra in Fig.1.  There are three stoichiometric positions
for oxygen in Bi-2212,  in atomic planes
containing Cu, the O(1) site; Sr planes, the O(2) site; and Bi
planes. The last oxygen atoms are
unobserved\cite{Tro90} owing to disorder in this plane. In addition
there is a small amount of non-stoichiometric
oxygen,  $\delta$-oxygen, too small to be observed directly by NMR
and whose location in the structure is not yet
established\cite{McE05}. The parts of the NMR spectra  in the middle
range of field in Fig. 1  are the central
transitions from the O(1) (broad) and O(2) (narrow) sites.  The
electronic coupling to
$^{17}$O is much stronger for O(1) compared with O(2) as is apparent
from the temperature dependences in Fig. 1.
This is confirmed by measurements of the spin-lattice relaxation rate
which are one order of magnitude larger for O(1)
as compared to O(2). The narrowness of the O(2) resonance indicates a
homogeneous electronic environment with
negligible spin-shift (Knight-shift) and uniform electric field
gradient.  For fast pulse repetition, as is the case for our
measurements in Fig.1, O(2) is partially saturated and comprises less
than 6\% of the spectrum at 60 K,
decreasing with decreasing temperature to  2\% at 40 K.  We  remove
it numerically in this range of temperature
and below this, it has negligible contribution to the spectrum.  The
O(2) resonance serves as a
useful marker for the zero spin-shift position of
$^{17}$O for both O(1) and O(2) sites.  In this report we focus on
$^{17}$O(1) NMR as a probe of the local
magnetic field in the CuO$_{2}$ plane.

\begin{figure}[h]
\vspace{0.3in}
\includegraphics[width=8 cm]{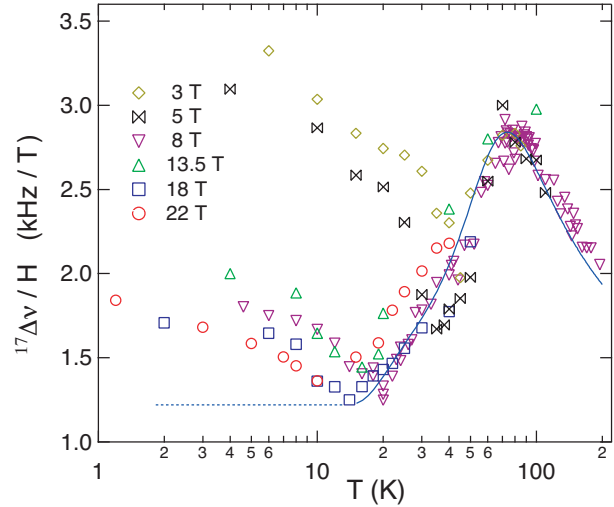}
\suppressfloats [b]
\caption[Linewidths]{Temperature dependence of the O(1) linewidth.
Temperature dependence of the linewidth of the
$^{17}$O NMR central transition presented as the square root of the
second moment of the frequency spectrum,
$^{17}\Delta\nu$, divided by magnetic field $H$ in the range 3 to 22
T. Fields above 8 T were obtained at the
National High Magnetic Field Laboratory in Tallahassee, Florida.
Measurements at 27 and 29 T are included in Fig. 3 and
4, but omitted here for clarity. On decreasing the temperature  there
is a common behavior independent of magnetic field,
represented by the smooth curve and its continuation to low
temperature (dotted curve), that can be understood in terms of
defects in the copper oxygen plane induced by oxygen doping. This
curve is given by, $\Delta \nu =  \Delta \nu_{0} +
KHD/T$, where
$\Delta
\nu_{0}$ is a background magnetic contribution to the linewidth
possibly associated with regions of the sample where
superconductivity is suppressed,
$K(T)$ is the measured Knight shift, $H/T$ is the
field-to-temperature ratio and $D$ is a Curie constant. The key
feature
of the data is the systematic break from this curve  that we identify
with the transition from a liquid to a solid vortex
state.}
\label{Fig2}
\end{figure}

Decreasing the temperature below the superconducting transition
temperature we find the
$^{17}$O(1) resonance peaks move to the left, {\it i.e.} to a higher
field at a fixed NMR frequency, (lower
frequency at fixed field) approaching the zero spin-shift position.
Simultaneously,  the central NMR line
narrows.  The decrease of the Knight shift to zero in the
superconducting state is a characteristic signature for
spin-singlet pairing.  The line broadening with decreasing
temperature in the normal state, Fig. 1 and 2, can be
associated with a Knight shift distribution introduced by the
$\delta$-oxygen.  Similar behavior has been
observed\cite{Bob97} for chemical impurities, like Ni, Zn, or
Li,  substituted for Cu in the CuO$_{2}$ plane in YBCO, and which
form a local moment. Their contribution to the $^{17}$O NMR
linewidth is given by a Curie law, proportional to the ratio of
applied magnetic field to temperature.   Additionally, we
find that below
$T_{c}$ this linewidth is proportional to the temperature dependent
Knight shift, which we measure independently,
thereby accounting for  the decrease
with decreasing temperature in Fig.2 as shown by the smooth curve,
assuming a temperature independent residual contribution
of 1.2 kHz/T. In this region, liquid vortex dynamics effectively
average to zero their associated local magnetic
fields.

On further cooling in the superconducting state there is a  sharp
onset for  a new contribution to the linewidth
which is not proportional to applied magnetic field.  The
well-defined temperature at which this line broadening
appears is plotted in Fig. 3, decreasing, but progressively more
slowly,  with increasing magnetic field in our
range
   between 3 and 29 Tesla.  We identify this behavior with the
formation of a solid vortex structure.  More
precisely, the extra linewidth corresponds to an inhomogeneous
magnetic field distribution  that is static on the
NMR time scale,
$\approx 0.1$ ms and is asymmetric; see the $T = 4$ K spectrum in
Fig. 1. This behavior is characteristic  of the
transition from liquid to solid vortex matter such as has been
observed by $\mu$SR\cite{Lee93} and
SANS\cite{Cub93} in Bi-2212,  and NMR in  YBCO\cite{Rey97}.
    Our observations are qualitatively consistent with extension to
high magnetic fields of  the vortex melting
phase diagram, $H < 0.1 $ T, determined by Hall
probe\cite{Kha96,Fuc98} and transport measurements\cite{Tor04}.
Khaykovich {\it et al.}\cite{Kha96} explore this behavior as a
function of oxygen doping (anisotropy).   NMR is
complementary to these methods with definite advantages for detecting
vortex melting at very high magnetic fields.  The
NMR or $\mu$SR spectrum is a direct map of the magnetic field
distribution from vortex supercurrents.  Specifically, NMR
with
$^{17}$O affords excellent resolution as a  magnetometer on the
atomic scale and has been exploited in previous
work\cite{Mit01} to spatially resolve and study excitations in the vortex core.

\begin{figure}[h]
\vspace{0.3in}
\includegraphics[width=8 cm]{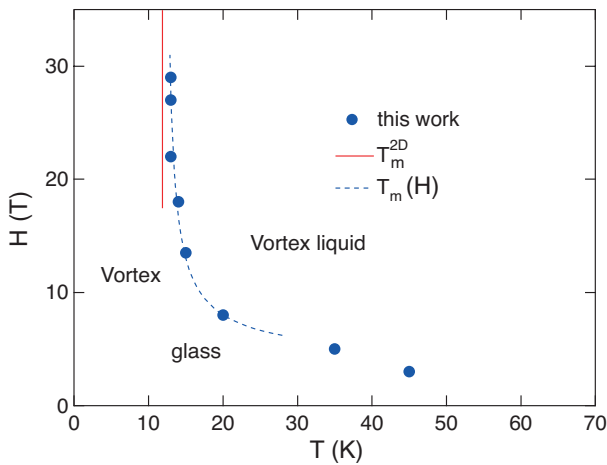}
\suppressfloats [b]
\caption[Phase Diagram]{The magnetic field-temperature phase diagram for
vortex melting in Bi-2212 for
$H||c$. The transition temperatures, T$_{m}$(H), are identified from the data,
as  shown in Fig. 2.  The two-dimensional vortex
melting transition temperature, T$_{m}^{2D}$ (vertical dashed line), is
determined  from a fit to the data. Deviations of the fit
from the data are  expected\cite{Gla91} for
$H \approx H_{cr}$}
\label{Fig3}
\end{figure}

The strong upward curvature of the phase diagram in Fig 3. has been
anticipated theoretically\cite{Gla91, Bla94}.  Torque
magnetometry in fields $H
\leq 5$ T gave similar indications\cite{Gam88}.  For highly
anisotropic superconductors the electromagnetic
interaction between vortices dominates the Josephson coupling between
planes.  In this high field limit the
simplest picture for vortex melting is a first order thermodynamic
transition\cite{Gla91}  given by
\begin{equation} T_{m}(H) = T_{m}^{2D}\left[ 1 +
\frac{b^{1/\nu}}{ln^{1/\nu}(H/H_{cr})}\right],
\end{equation} for $H$ larger than a crossover field,
\begin{equation} H_{cr} \approx \frac{2\pi
\phi_{0}}{\gamma^{2}d^{2}}\:ln\left( \frac{\gamma d}{\xi_{ab}}
\sqrt{1 + \frac{(4\pi
\lambda_{c})^{2}d} {2(\phi_{0}\xi_{ab})^{2}}k_{B}T}\:\:\right).
\end{equation} The limiting two-dimensional melting temperature in Eq. 1. is,
\begin{equation} T_{m}^{2D} = A
\frac{\phi_{0}^{2}d} {8\pi \sqrt{3} k_{B}(4\pi\lambda_{ab})^{2}}.
\end{equation} where $0.605 \leq A \leq 0.615$, comes from numerical
calculations (Koshelev, A.E., private
communication).  This spread in $A$ reflects the existence of an
intermediate phase since the ideal
   two-dimensional melting scenario proceeds in two steps with an
intervening hexatic phase bounded by continuous
transitions\cite{Bla94}.  The melting temperature, Eq. 3, depends on
the layer spacing\cite{Kog93}
$d = 1.5$ nm, the flux quantum $\phi_{0}$, Boltzmann's constant
$k_{B}$, and the single superconductive parameter
$\lambda_{ab}$, which is the penetration depth for supercurrents in
the CuO$_{2}$ plane. The crossover field,
$H_{cr}$, depends mainly on the product of
$d$ and the mass anisotropy factor, $\gamma = \lambda_{c}/\lambda_{ab} =
\xi_{ab}/\xi_{c}$.  Out-of plane components are denoted by a
subscript c and $\xi$ is the coherence length. The
numerical constants in the theory, Eq. 1. are
$b \approx 1$ and the exponent $\nu = 0.37$.

\begin{figure}[h]
\vspace{0.3in}
\includegraphics[width=8 cm]{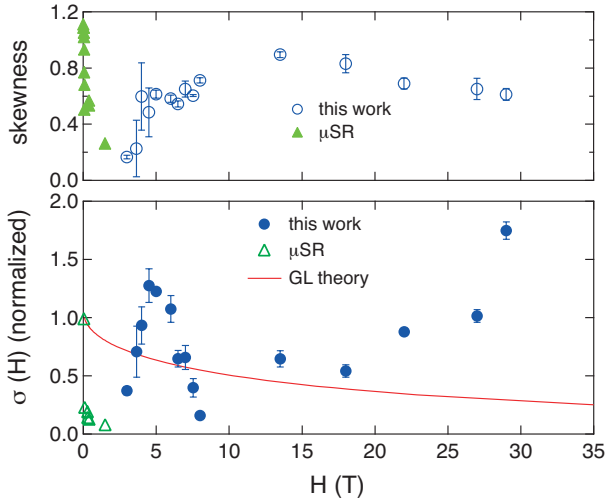}
\suppressfloats [b]
\caption[LT-Linewidth]{Magnetic field dependence of the vortex linewidth and
asymmetry.  The vortex contribution to the second
moment of the NMR spectra, $\sigma$(H), at low temperature, $T \approx 4$ K, is
given in the lower panel as the difference between the
data shown in Fig. 3 and the background contribution,
   smooth curves in Fig. 2, expressed as a second moment.  A comparison
with that expected from rigid-line vortices
is calculated from  Ginzburg-Landau theory\cite{Bra03}, smooth curve,
taking the upper  critical field to be 230
T.  Low field $\mu$SR measurements\cite{Inu93, Aeg98} up to 1.5 T are
shown as triangles with the lowest field less than H*
in the Bragg glass phase, consistent with three dimensional ordering
there, but dropping dramatically in the vortex glass
phase.  The linewidth data are normalized to the theoretical value at low
field. In the upper panel we show the skewness,
$\alpha = \langle \left[\nu - \nu_{1}\right]^{3}  \rangle^{1/3} /
\langle \left[\nu - \nu_{1}\right]^{2}
\rangle^{1/2} $ where $\nu_{1}$ is the first moment of the spectrum
and the average is over the spectrum. A decrease in
the second moment $H \approx 6$ T with no drop in asymmetry suggests
intra-planar vortex ordering. The error bars are statistical, calculated from
the NMR spectra.}
\label{Fig4}
\end{figure}

The NMR vortex lineshape is asymmetric but less so than for a perfect
line-vortex lattice suggesting that the vortex structure
is somewhat disordered.  In fact, $\mu$SR and SANS show that at low
temperature and $H > 0.1$ T, it is a vortex glass.  We
will assume that the difference in energy between disordered  and
perfect vortex structures, at least in high magnetic
field, is small compared to the  energy for condensation from liquid
to solid, and we use the framework of Eq. 1-3. to
analyze the freezing that we have observed.  From a fit to the data,
dotted curve in Fig. 3, we find the cross-over field
$H_{cr} = 2.5$ T and $\gamma \approx 78$ and the two-dimensional
melting transition temperature
\mbox{$T_{m}^{2D}= 12 \pm 1$ K} for which the penetration depth is
$\lambda_{ab} = 220 \pm 10$ nm.  Precision
measurements of the absolute value of the penetration depth at low
temperature are notoriously difficult.  From
earlier reports for
   Bi-2212, $\lambda_{ab}$ is  $\approx 210$ nm from magnetization
data\cite{Kog93}, $269 \pm 15$ nm from cavity
resonance methods\cite{Pro00} and
   $\approx 180$ nm from $\mu$SR\cite{Lee93}.  These results are in
agreement with what we report here.
 From  torque
measurements, Iye
{\it et al.}\cite{Iye92}   found
$\gamma \geq 200$ for nearly optimum doped samples, and Watauchi {\it et
al.}\cite{Wat01} obtained 91  on overdoped
material  using resistivity methods.  Although
$\gamma \approx 78$ is in this range it must be considered only
approximate since the theory is imprecise in the crossover
field region.  If we constrain $\gamma$, varying it over this wide
range, our fit for $T_{m}^{2D}$ is unaffected.  For a
less overdoped crystal,
$T_{c} = 85$ K, we find a slightly higher value, $T_{m}^{2D} = 16$ K.

Experiments consistent with the two-dimensional melting theory have
basic significance.  But they are also
relevant to applications of superconductivity at very high magnetic
field.   One of the most promising materials
for magnet wire\cite{Has04,Sca04} to make a  30 T, superconducting, NMR magnet
\cite{COHMAG} is Bi-2212. To achieve this goal there will be an upper
limit on the operational temperature of the
magnet determined by critical currents that are more easily
stabilized in the vortex solid phase.

At low magnetic fields, $H << H_{cr}$,  it is well-established that
there is a complex phase diagram for Bi-2212
with transitions on cooling from liquid vortex matter to a solid.  At
low temperature an increasing magnetic field drives a
transition at $H^{*} \approx 0.1$ T from  Bragg glass\cite{Kle01} to
vortex glass, indicated by an abrupt increase
in the symmetry of the $\mu$SR spectrum\cite{Lee93}, a decrease in
its second moment\cite{Lee93}, and disappearance of Bragg
peaks in SANS\cite{Cub93}.  At this transition vortices lose
coherence between planes. The resulting  destructive
interference between vortices on adjacent planes  averages out the
magnetic field distribution\cite{Inu93, Aeg98, Bra91}
that one might expect for straightline vortices such as are seen in
the Bragg glass phase\cite{Lee93}.  This interference
reduces both asymmetry and linewidth in the $\mu$SR spectrum.  With
even larger fields  the second moment of the
$^{17}$O NMR spectrum,
$\sigma(H)$, ($\approx$ linewidth squared) progressively increases
and asymmetry is restored,
until the spectrum linewidth collapses at 5 T.   There are two
possibilities for this anomalous behavior at 5 T.
Either the  transition is field-induced-ordering (ordering in-plane)
or field-induced-disordering (further disorder
between planes).  The significant asymmetry in the spectrum in the
high field phase compared to that at our lowest fields
favors vortex ordering with increasing field.   This scenario might follow if
interplanar coupling is weakened with increasing field compared to
intraplanar interactions,  reducing
frustration that originated from vortices in adjacent planes.  As a consequence
vortices order two-dimensionally and we observe the second moment
decrease abruptly.  At much larger fields, we can
see that the two-dimensional vortex solid becomes progressively more
disordered. On theoretical grounds\cite{Lar79}
arbitrarily small amounts of impurity will lead to disorder of the
vortex lattice.  Sensitivity to quenched-in disorder
should increase progressively with increasing magnetic field as the
lattice spacing decreases with fixed vortex size.

   To make a comparison with straight line vortices
   we have performed a Ginzburg-Landau calculation\cite{Bra03} of the
corresponding magnetic field distribution,
Fig. 4. Our measurement is of a similar magnitude as expected from this
calculation, although the GL approximation does not capture the
details of the observed field dependence.   In fact
the transition from liquid to a solid  in high magnetic fields is
predicted to occur first to a supersolid phase
and then at lower temperatures to a decoupled solid phase where
defects in the vortex lattice become
zero-dimensional\cite{Bla94,Cra97}.  NMR techniques  may be helpful
in exploring these fascinating aspects of
vortex behavior in strongly anisotropic superconductors.

\hfill\break

{\bf Acknowledgements}  We gratefully acknowledge discussions with
L.I. Glazman, A.E.
Koshelev, D.C. Larbalastier, J.A. Sauls, and E. Zeldov and
contributions  from E.E. Sigmund and P. Sengupta. We are grateful to Robert
Smith for a detailed study of the satellites of O(1) in the vortex solid
state. This work was supported by the Department of
Energy, contract DE-FG02-05ER46248 and the National  High Magnetic Field
Laboratory, the National Science Foundation, and the State of
Florida.

Experimental work and analysis was performed by BC, WPH, VFM, APR,
and PLK.  Samples were prepared by PG and DGH.\hfill\break

{\bf Author Information}   Correspondence should be addressed to
W.P.H. (w-halperin@northwestern.edu).

\end{document}